\documentclass[twocolumn,preprintnumbers,superscriptaddress,amsmath,amssymb, bibnotes,nofootinbib]{revtex4}
\usepackage{amsthm}
\usepackage{amsmath}
\usepackage{amssymb}
\usepackage{graphicx}
\usepackage{url}
\usepackage{dsfont}
\usepackage{float}
\usepackage{bm}
\usepackage{ifthen}
\usepackage[usenames,dvipsnames]{color}
\usepackage{mathrsfs}
\usepackage[colorlinks=true,citecolor=blue,urlcolor=black]{hyperref}
\usepackage{float}
\usepackage{afterpage}
\usepackage{subfigure}
\usepackage{adjustbox}
\usepackage{times}

\newcommand{\be}{\begin{equation}}
\newcommand{\ee}{\end{equation}}


\newcommand{\sket}[1]{{\ensuremath{\lvert#1\rangle}}}
\newcommand{\lket}[1]{{\ensuremath{\left\lvert#1\right\rangle}}}
\newcommand{\ket}[1]{\if@display\lket{#1}\else\sket{#1}\fi}
\newcommand{\sbra}[1]{{\ensuremath{\langle#1\rvert}}}
\newcommand{\lbra}[1]{{\ensuremath{\left\langle#1\right\rvert}}}
\newcommand{\bra}[1]{\if@display\lbra{#1}\else\sbra{#1}\fi}
\newcommand{\sbraket}[2]{{\ensuremath{\langle#1\rvert#2\rangle}}}
\newcommand{\lbraket}[2]{{\ensuremath{\left\langle#1\!\left\rvert\vphantom{#1}#2\right.\!\right\rangle}}}
\newcommand{\braket}[2]{\if@display\lbraket{#1}{#2}\else\sbraket{#1}{#2}\fi}

\newcommand{\sketbra}[2]{{\ensuremath{\lvert #1\rangle\!\langle #2\rvert}}}
\newcommand{\lketbra}[2]{{\ensuremath{\left\lvert #1\right\rangle\!\!\left\langle #2\right\rvert}}}
\newcommand{\ketbra}[2]{\if@display\lketbra{#1}{#2}\else\sketbra{#1}{#2}\fi}

\theoremstyle{plain}

\theoremstyle{definition}

\begin{document}
\title{Discrete-variable measurement-device-independent quantum key distribution suitable for metropolitan networks}

\author{Feihu Xu}
\email{fhxu@mit.edu}
\affiliation{Research Laboratory of Electronics, Massachusetts Institute of Technology,
77 Massachusetts Avenue, Cambridge, Massachusetts 02139, USA}
\author{Marcos Curty}
\affiliation{EI Telecomunicaci\'on, Dept. of Signal Theory and Communications, University of Vigo, E-36310 Vigo, Spain}
\author{Bing Qi}
\affiliation{Quantum Information Science Group, Computational Sciences and Engineering Division,
Oak Ridge National Laboratory, Oak Ridge, TN 37831-6418, USA}
\author{Li Qian}
\affiliation{Center for Quantum Information and Quantum Control, Dept. of Electrical \& Computer Engineering and Dept. of Physics,
University of Toronto, Toronto, Ontario, M5S 3G4, Canada}
\author{Hoi-Kwong Lo}
\affiliation{Center for Quantum Information and Quantum Control, Dept. of Electrical \& Computer Engineering and Dept. of Physics,
University of Toronto, Toronto, Ontario, M5S 3G4, Canada}

\date{\today}
\begin{abstract}
We demonstrate that, with a fair comparison, the secret key rate of
discrete-variable measurement-device-independent quantum key distribution
(DV-MDI-QKD) with high-efficiency single-photon detectors and good 
system alignment is typically rather high and thus highly suitable for 
not only long distance
communication but also metropolitan networks. The previous reservation 
on the key rate and suitability of DV-MDI-QKD for metropolitan networks 
expressed by Pirandola et al. [1] was based on an unfair comparison with low-efficiency
detectors and high quantum bit error rate, and is, in our opinion, unjustified.
\end{abstract}

\maketitle

In a recent Article in {\it Nature Photonics}, Pirandola {\it et al.}~\cite{CVMDI} claim that the achievable secret key rates of discrete-variable measurement-device-independent quantum key distribution
(DV-MDI-QKD)~\cite{MDIQKD} are ``typically very low, unsuitable for the demands of a metropolitan network'' and introduce a continuous-variable (CV) MDI-QKD protocol capable of providing key rates which, they claim, are
``three orders of magnitude higher'' than those of DV-MDI-QKD. We believe, however, that such statements are
unjustified as they are based on an unfair comparison between the two platforms.
Here, we show that, with a fair comparison, the secret key rate of DV-MDI-QKD with high-efficiency
single-photon detectors (SPDs) and good system alignment is typically rather high and thus highly suitable for
not only long-distance communication but also
metropolitan networks. The claimed very low key rate of DV-MDI-QKD in~\cite{CVMDI} was due to a combination
of pessimistic assumptions of low-efficiency SPDs and rather high quantum bit error rate (QBER).

It is well-known that CV-QKD could offer higher key rates than DV-QKD
at relatively short distances~\cite{CVQKD1}, while the exact enhancement factor depends on technology.
In their work, Pirandola {\it et al.}~\cite{CVMDI} consider
nearly perfect devices for CV-MDI-QKD (i.e., a relay with overall detection efficiency $\eta=98$\% and an excess noise $\varepsilon\simeq0.01$), but, rather surprisingly, use
low-performance off-the-shelf fiber-optical components with $\eta=14.5$\% and QBER $=2.94$\% (which corresponds to the misalignment error of $1.5$\% assumed in~\cite{MDIQKD})
for DV-MDI-QKD.
Such comparison is unfair as it ignores the existence of high-efficiency  SPDs ($\eta=93$\%~\cite{SSPD}) and much lower achieved QBER ($0.25$\%~\cite{MDIQKDexp2}),
which could easily be used in DV-MDI-QKD. In addition, they
evaluate the most favourable case for CV-MDI-QKD with a
relay located very close to the sender (see Fig.~5d in ref.~\cite{CVMDI}),
which is not necessarily the case in a real network.
In doing so, it is not surprising that the enhancement factor is noteworthy.

Here, we would like to point out that the situation changes significantly with better SPDs~\cite{SSPD} and lower QBER \cite{MDIQKDexp2} in DV-MDI-QKD. For CV-MDI-QKD, we use the same optimistic assumptions on the experimental parameters employed
in ref.~\cite{CVMDI}. We first consider the highly asymmetric case where the relay is placed very close to
one of the users, Alice, so that there is no channel loss between Alice and the relay.
Figs.~\ref{Fig}a and~\ref{Fig}b illustrate this scenario
in the asymptotic limit of an infinitely long key (see Appendices for further details).
In this case, at a typical metropolitan distance (say $20$ km of standard telecom fiber of loss $0.2$ dB/km
used as an example in~\cite{CVMDI}),
the key rate of DV-MDI-QKD is about $0.02$ bits/use, which is actually quite high
(i.e., approximately two orders of
magnitude away from the fundamental limit~\cite{takeoka})
and suitable for metropolitan networks.
Moreover, this result is only slightly lower but comparable to that of CV-MDI-QKD.
Notice that, even in this asymmetric case, the advantage in key rate of CV-MDI-QKD is less than
one order of magnitude for total system loss beyond about only $2.5$ dB.

In a general network, it is quite likely that a relay is far away from both users. In the symmetric case where the relay is placed in the middle between them, we see that DV-MDI-QKD compares favorably with CV-MDI-QKD. This is illustrated in Figs.~\ref{Fig}c and~\ref{Fig}d (see Appendices for further details). 
As shown there, while the key rate of DV-MDI-QKD is still about $0.02$ bits/use at 20 km of telecom fiber, the key rate of CV-MDI-QKD drops to zero at already about $6.25$ km.
Indeed, due to optical fiber loss the maximum distance of CV-MDI-QKD is already limited to about 7.6 km (1.52 dB)~\cite{CVMDI}. Note that the performance of a network with arbitrary configuration will be between the asymmetric and the symmetric cases. 

\begin{figure*}[htb]
\centering
{\includegraphics [width=13.5cm]{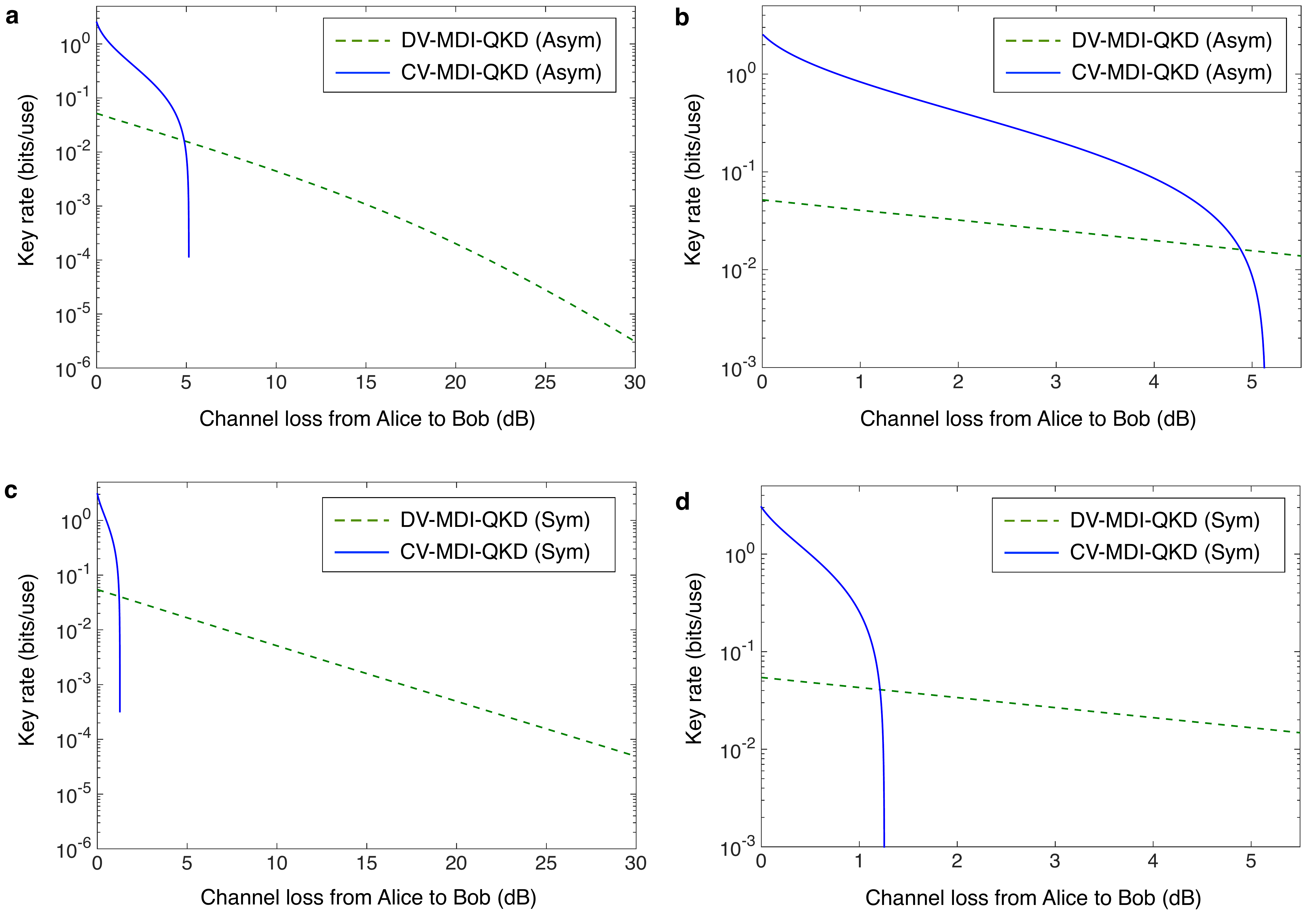}}
\caption{Secret key rate of DV-MDI-QKD (green dashed line) versus CV-MDI-QKD (blue solid line).
For the former we consider $\eta=93$\%~\cite{SSPD}, QBER$=0.25$\%~\cite{MDIQKDexp2}, and an error correction inefficiency factor of $1.16$. For the later
we use experimental parameters from ref.~\cite{CVMDI} (i.e., $\eta=98$\%, $\varepsilon\simeq0.01$, and reconciliation efficiency $\xi=0.97$).
\textbf{a},
Asymmetric scenario with a relay located close to Alice. The solid blue line corresponds to Fig.~5d(i) in ref.~\cite{CVMDI}. We see that the advantage in key rate of CV-MDI-QKD is less than
one order of magnitude for total system loss beyond about only $2.5$ dB. \textbf{b},
A zoom-in of \textbf{a} for the low loss regime.
\textbf{c}, Symmetric scenario with a relay placed in the middle of the users. This result seems to indicate that CV-MDI-QKD is unsuitable for applications in a symmetric case as the
key rate drops to zero at already about 1.25 dB.
Indeed, due to optical fiber loss the maximum distance of CV-MDI-QKD is already limited to about $7.6$ km (1.52 dB)~\cite{CVMDI}.
\textbf{d}, As in \textbf{b}, but now referred to \textbf{c}.} \label{Fig}
\end{figure*}

It is worthy to mention as well several practical challenges associated with CV-MDI-QKD. First, the outstanding experimental data from ref.~\cite{CVMDI} comes from a proof-of-principle demonstration where Alice and Bob are very close to each other, thus bypassing one of the major experimental challenges in CV-MDI-QKD implementations---establishing a reliable phase reference between two spatially separated users. Note that the Bell state measurement of DV-MDI-QKD is insensitive to phase noise~\cite{MDIQKDexp2}, thus much easier to implement. Second, coupling losses and state-of-the-art homodyne detectors could render it difficult to achieve $\eta=98$\% for fiber-based CV-MDI-QKD communications at telecom wavelengths (e.g., $\eta$ is about 60\% in ref.~\cite{CVQKD}). Third, in the realistic finite-key length regime, current composable security proofs against general attacks for CV-QKD with coherent states seem to fall short on providing useful finite-size key estimates~\cite{Leverrier}. This strongly contrasts with DV-MDI-QKD~\cite{finite_mdiQKD}. Finally, if one considers the key rate per second, currently the high repetition rate of DV-QKD (i.e., 1 GHz~\cite{shields}) versus CV-QKD (i.e., 1 MHz~\cite{CVQKD}) means that DV-QKD has an advantage in this regard.

All these factors combine to lead us to conclude that, contrary to the claims of
Pirandola {\it et al.}~\cite{CVMDI},
DV-MDI-QKD has a high key rate and is highly versatile and suitable for
most metropolitan network configurations. Moreover, experimental DV-MDI-QKD has already been done even at a distance of $200$ km over telecom fibers~\cite{MDIQKDexp2}.
CV-MDI-QKD has the potential for applications in some (e.g., highly asymmetric)
metropolitan network configurations, but fares poorly in a symmetric network setting when a relay
is far away from both users. Furthermore, a number of challenges including finite-key analysis with composable security, 
establishing a reliable phase reference between two remote users and low repetition rate need to be overcome before CV-MDI-QKD can be securely deployed in practice.

\section*{Acknowledgements}
The authors thank 
V. Anant, E. Diamanti, A. Leverrier, Z.-Y.~Li, J.~Shapiro, Y. Shi, Z. Yuan, Q. Zhang and Z. Zhang 
for valuable discussions. Support from the Office of Naval Research (ONR), the Air Force Office of Scientific Research (AFOSR),
the Galician Regional Government (program ``Ayudas para proyectos de investigaci\'on desarrollados por investigadores emergentes'', and consolidation of Research Units: AtlantTIC), the 
Spanish Ministry of Economy and Competitiveness (MINECO), the ``Fondo Europeo de Desarrollo Regional'' (FEDER) through grant TEC2014-54898-R, NSERC, CFI, and ORF is gratefully acknowledged.

\appendix

\section{DV-MDI-QKD}

Here, we present the detailed models that we use for the numerical simulations of DV-MDI-QKD shown in Fig.~\ref{Fig}.

The secure key rate of decoy-state DV-MDI-QKD in the asymptotic limit of an infinitely long key is given by~\cite{MDIQKD}
\begin{equation} \label{Eqn:Key:DV}
\begin{aligned}
    R_{\text{DV}} = p_{11}^{\rm Z}Y_{11}^{\rm Z}[1-H_{2}(e^{\rm X}_{11})]
    -Q^{\rm Z}f_{e}(E^{\rm Z})H_{2}(E^{\rm Z}),
\end{aligned}
\end{equation}
where $p_{11}^{\rm Z}=\mu_{\rm A}\mu_{\rm B} e^{-(\mu_{\rm A}+\mu_{\rm B})}$ denotes the joint probability that both Alice and Bob generate a single-photon pulse, and with
$\mu_{\rm A}$ and $\mu_{\rm B}$ being, respectively, the intensity of Alice and Bob's signal states; the parameters $Y_{11}^{\rm Z}$ and $e^{\rm X}_{11}$ are, respectively, the yield in the rectilinear ($\rm Z$) basis and the error rate in the diagonal ($\rm X$) basis, given that both Alice and Bob send single-photon states; $H_{2}(x)=-x\log_2{(x)}-(1-x)\log_2{(1-x)}$ is the binary Shannon entropy function; the terms $Q^{\rm Z}$ and $E^{\rm Z}$ denote, respectively, the overall gain and quantum bit error rate (QBER) in the $\rm Z$ basis when both Alice and Bob emit a signal state; and $f_{e}(E^{\rm Z})\geq 1$ is the error correction inefficiency function.

The quantities $Q^{\rm Z}$ and $E^{Z}$ are directly measured in the experiment, while $Y_{11}^{\rm Z}$ and $e^{\rm X}_{11}$ can be estimated using the decoy-state method~\cite{decoyMDI}.
Importantly, it has been shown that the use of two decoy states is already enough to obtain a tight estimation for $Y_{11}^{\rm Z}$ and $e^{\rm X}_{11}$~\cite{decoyMDI2}.

To model experimental errors, we employ the method proposed in ref.~\cite{decoyMDI} for polarisation encoding DV-MDI-QKD~\cite{PolExp1,PolExp2}. See also refs.~\cite{timebin1,timebin2} for alternative models
suitable for time-bin encoding systems~\cite{timebinexp1,timebinexp2}. In particular, we use two unitary operators, located at the input arms of the beamsplitter within the relay (see Appendix~B in ref.~\cite{decoyMDI}),
to simulate the
intrinsic error rate, denoted as $e_{\rm d}$, due to the misalignment and instability of the optical system. In addition,
we consider threshold SPDs with detection efficiency $\eta_{\rm d}$ and dark count rate $Y_{0}$. Furthermore, for simplicity,
we consider the asymptotic case where Alice and Bob use an infinite number of decoy states. As already mentioned, the practical situation with a
finite number of decoy settings provides similar results~\cite{decoyMDI2}.
In this scenario, we have that the parameters $Y_{11}^{\rm Z}$ and $e^{\rm X}_{11}$ have the form~\cite{decoyMDI}
\begin{eqnarray} \label{Q11e11:nonideal:final} \nonumber
    Y_{11}^{\rm Z} &=& (1-Y_{0})^{2}\Bigg[4Y_{0}^{2}(1-\eta_{\rm A}\eta_{\rm d})(1-\eta_{\rm B}\eta_{\rm d})\\ \nonumber
               &+&2Y_{0}\Big(\eta_{\rm A}\eta_{\rm d}+\eta_{\rm B}\eta_{\rm d}-\frac{3}{2}\eta_{\rm A}\eta_{\rm B}\eta_{\rm d}^{2}\Big)+\frac{1}{2}\eta_{\rm A}\eta_{\rm B}\eta_{\rm d}^{2}\Bigg],\\
    e^{\rm X}_{11}&=&\frac{1}{2}-\ \frac{(1-Y_{0})^{2}\eta_{\rm A}\eta_{\rm B}\eta_{\rm d}^{2}(1-e_{\rm d})^{2}}{4Y_{11}^{\rm X}}, \\ \nonumber
\end{eqnarray}
where $\eta_{\rm A}$ ($\eta_{\rm B}$) denotes the channel transmittance from Alice (Bob) to the relay. That is,
$\eta_{\rm A}=10^{-\alpha l_{\rm A}/10}$, where $\alpha$ is the loss coefficient of the channel that connects Alice with the relay measured in dB/km, and $l_{\rm A}$ is the length of this channel measured in km.
The definition of $\eta_{\rm B}$ is analogous. Moreover, we have that $Y^{\rm X}_{11}=Y^{\rm Z}_{11}$.

For simulation purposes only, we use the value of $Q^{\rm Z}$ and $E^{\rm Z}$ provided in Appendix~B of ref.~\cite{decoyMDI}. For completeness, 
we include the mathematical expressions below. In particular, 
\begin{eqnarray}
Q^{\rm Z}&=&\frac{1}{2}\left(\Omega_1+\Omega_2\right),\\ \nonumber
E^{\rm Z}&=&\frac{\Omega_1}{\Omega_1+\Omega_2},
\end{eqnarray}
where the parameters $\Omega_1$ and $\Omega_2$ are given by
\begin{eqnarray}
\Omega_1&=&2e^{-\frac{\gamma}{2}}(1-Y_{0})^{2}\Big[I_{0}(\beta)+I_{0}(\beta-2\beta e_{\rm d})\\ \nonumber
&+&2(1-Y_{0})^{2}e^{-\frac{\gamma}{2}}-2(1-Y_{0})e^{-\frac{\gamma(1-e_{\rm d})}{2}}I_{0}(e_{\rm d}\beta)   \\ \nonumber
&-&2(1-Y_{0})e^{-\frac{\gamma e_{\rm d}}{2}}I_{0}(\beta-e_{\rm d}\beta)\Big], \\ \nonumber
\Omega_2&=&2e^{-\frac{\gamma}{2}}(1-Y_{0})^{2}\Big[1+I_{0}(2\lambda)+2(1-Y_{0})^{2}e^{-\frac{\gamma}{2}} \\ \nonumber
&-&2(1-Y_{0})e^{-\frac{\omega}{2}}I_{0}(\lambda)-2(1-Y_{0})e^{-\frac{\gamma-\omega}{2}}I_{0}(\lambda)\Big],
\end{eqnarray}
with $I_{0}(x)$ being the modified Bessel function, and where
\begin{eqnarray}
\gamma&=&(\mu_{\rm A}\eta_{\rm A}+\mu_{\rm B}\eta_{\rm B})\eta_{\rm d}, \\ \nonumber
\beta&=&\eta_{\rm d}\sqrt{\mu_{\rm A}\mu_{\rm B}\eta_{\rm A}\eta_{\rm B}}, \\ \nonumber
\lambda&=&\beta\sqrt{e_{\rm d}(1-e_{\rm d})}, \\ \nonumber
\omega&=&\mu_{\rm A}\eta_{\rm A}\eta_{\rm d}+e_{\rm d}(\mu_{\rm B}\eta_{\rm B}-\mu_{\rm A}\eta_{\rm A})\eta_{\rm d}.
\end{eqnarray}

\begin{table}[hbt]
\centering
\begin{tabular}{c @{\hspace{0.5cm}} c @{\hspace{0.5cm}} c @{\hspace{0.5cm}} c} \hline
$\eta_{\rm d}$ & $e_{\rm d}$ & $Y_{0}$ & $f_{e}(E^{\rm Z})$ \\
\hline
93\%~\cite{SSPD} & 0.1\%~\cite{MDIQKDexp2} & $10^{-6}$~\cite{SSPD}  & 1.16~\cite{corr} \\
\hline
\end{tabular}
\caption{Experimental parameters considered in DV-MDI-QKD.} \label{exp_dv}
\end{table}
In our simulation, we employ the experimental parameters shown in Table~\ref{exp_dv}. That is, 
we consider high-efficiency WSi superconducting nanowire single-photon
detectors (SNSPDs) with $\eta_{\rm d}=93\%$ and $Y_0=10^{-6}$ (per pulse)~\cite{SSPD}, and assume an intrinsic error rate $e_{\rm d}=0.1\%$ (which corresponds to the QBER of 0.25\% obtained in the 200 km DV-MDI-QKD experiment reported in ref.~\cite{MDIQKDexp2}). In addition, we use $f_{e}(E^{\rm Z})=1.16$~\cite{corr}.
The resulting lower bound on the secret key rate $R_{\text{DV}}$ given by Eq.~(\ref{Eqn:Key:DV}) is illustrated in Fig.~\ref{Fig} in the main text, where we have numerically optimised
the values of the intensities $\mu_{\rm A}$ and $\mu_{\rm B}$. That is, for a given total system loss, 
we use a Monte Carlo simulation method  to select the value of $\mu_{\rm A}$ and $\mu_{\rm B}$ such that $R_{\text{DV}}$ is maximum.

To conclude this section, let us emphasise that the high-efficiency SNSPDs reported in ref.~\cite{SSPD} have been already successfully applied in various recent QKD demonstrations such as high-dimensional
QKD~\cite{HDQKD,HDQKD2} and a proof-of-principle demonstration of DV-MDI-QKD~\cite{SSPDMDI}.

\section{CV-MDI-QKD}\label{appB}

In this Appendix we present the detailed models that we use for the numerical simulations of CV-MDI-QKD shown in Fig.~\ref{Fig}.

To evaluate the secure key rate of CV-MDI-QKD we
follow Section~E from the Supplementary Information of ref.~\cite{CVMDI}. 
We have that the general expression of the key rate formula has the form
\begin{equation} \label{Eqn:Key:CV}
\begin{aligned}
    R_{\text{CV}} = \xi I_{\text{AB}} - I_{\text{E}},
\end{aligned}
\end{equation}
where $\xi$ is the reconciliation efficiency of the error correction code,
and $I_{\text{AB}}$ and $I_{\text{E}}$ denote, respectively, Alice and Bob's mutual information and Eve's stolen information on Alice's key.

To derive an explicit formula for the secret key rate given by Eq.~(\ref{Eqn:Key:CV}), 
Pirandola {\it et al.}~\cite{CVMDI} consider a ``realistic Gaussian attack against the two links'', which definitively 
provides an upper bound on the secure key rate. Here, we assume the most favourable situation for 
CV-MDI-QKD by considering that such realistic Gaussian attack is indeed optimal and the resulting key rate is 
achievable. In doing so, we have that the quantity $I_{\text{AB}}$ can be expressed as~\cite{CVMDI}
\begin{eqnarray} \label{Eqn:IAB}
I_{\text{AB}}=\log_2 \left(\frac{\phi+1}{\chi}\right),
\end{eqnarray}
where $\phi$ is the modulation variance in shot-noise units and $\chi$ represents the so-called equivalent noise.

Moreover, for simplicity, we consider the key rate in the limit of large modulation ($\phi\gg 1$).
In this scenario,
and considering first the asymmetric case $\eta_{\rm A}\neq \eta_{\rm B}$, we have that $I_{\text{E}}$ is given by~\cite{CVMDI}
\begin{eqnarray} \label{Eqn:IE}
    I_{\text{E}}&=&h(\beta)+\log_2(\gamma)-h(\delta),
\end{eqnarray}
where the parameters $\beta$, $\gamma$ and $\delta$ have the form
\begin{eqnarray} \label{Eqn:IE2}
    \beta&=&\frac{\eta_{\rm A}\eta_{\rm B}\chi-(\eta_{\rm A}+\eta_{\rm B})^2}{|\eta_{\rm A}-\eta_{\rm B}|(\eta_{\rm A}+\eta_{\rm B})}\nonumber\\
    \gamma&=&\frac{e|\eta_{\rm A}-\eta_{\rm B}|(\phi+1)}{2(\eta_{\rm A}+\eta_{\rm B})} \nonumber\\
   \delta&=&\frac{\eta_{\rm A}\chi-(\eta_{\rm A}+\eta_{\rm B})}{\eta_{\rm A}+\eta_{\rm B}}.
\end{eqnarray}
Here, the term $e$ denotes Euler's number, and the equivalent noise $\chi$ is given by
\begin{eqnarray} \label{Eqn:chi}
\chi=\frac{2(\eta_{\rm A}+\eta_{\rm B})}{\eta_{\rm A}\eta_{\rm B}\eta_{\rm d}}+\varepsilon,
\end{eqnarray}
with $\varepsilon$ being the excess noise. Finally, the function $h(x)$, which appears in 
Eq.~(\ref{Eqn:IE}),
has the form
$$
h(x)=\left(\frac{x+1}{2}\right)\log_2\left(\frac{x+1}{2}\right)-\left(\frac{x-1}{2}\right)\log_2\left(\frac{x-1}{2}\right).
$$

In the symmetric case where $\eta_{\rm A}=\eta_{\rm B}=\eta$, we have that $I_{\text{E}}$ is given by~\cite{CVMDI}
\begin{eqnarray} \label{Eqn:IEsym}
    I_{\text{E}} = \log_2\left(\frac{e^2(\chi-4)(\phi+1)}{16}\right)-h\left(\frac{\chi}{2}-1\right),
\end{eqnarray}
where the equivalent noise $\chi$ has now the form
\begin{eqnarray} \label{Eqn:chi2}
\chi=\frac{4}{\eta\eta_{\rm d}}+\varepsilon.
\end{eqnarray}

In the simulation shown in Fig.~\ref{Fig} in the main text, we consider the same experimental parameters used in ref.~\cite{CVMDI}. 
They are illustrated in Table~\ref{exp_cv}.
\begin{table}[hbt]
\centering
\begin{tabular}{c @{\hspace{0.5cm}} c @{\hspace{0.5cm}} c @{\hspace{0.5cm}} c} \hline
$\eta_{\rm d}$ & $\varepsilon$ & $\phi$ & $\xi$ \\
\hline
$98$\% & $0.01$ & $60$  & 0.97 \\
\hline
\end{tabular}
\caption{Experimental parameters considered in CV-MDI-QKD~\cite{CVMDI}.} \label{exp_cv}
\end{table}

\section{TGM bound}

The secret key rates obtained in the previous sections can be compared with the fundamental upper bound (per optical mode) for coherent-state QKD provided in ref.~\cite{takeoka} (so-called
TGW bound). This bound has the form
\begin{equation}
R_{\text{TGW}} = \log_2\left(\frac{1+\eta_{\rm A}\eta_{\rm B}}{1-\eta_{\rm A}\eta_{\rm B}}\right).
\end{equation}
It can be shown, for instance, that at a typical metropolitan distance (say 20 km of standard telecom fiber of loss 0.2 dB/km used as an example in~\cite{CVMDI}), 
the key rate of DV-MDI-QKD is approximately two orders of magnitude away from this fundamental limit.

\section{Source requirements in CV-MDI-QKD}\label{appD}

Here, we briefly discuss challenges associated with the state preparation process in CV-MDI-QKD. 

As already mentioned in the main text, the experimental configuration using one laser feeding two closely spaced Alice and Bob on the optical table bypasses one of the major experimental challenges in CV-MDI-QKD, namely, establishing a reliable phase reference between two remote users. Even if a common laser is used for both Alice and Bob, distributing phase-stable signals over fibre to two remote sites can still be a practical challenge. 

Furthermore, it is important to emphasise that one fundamental assumption in MDI-QKD is that Eve cannot interfere with Alice and Bob's
state preparation processes~\cite{MDIQKD}. To justify the above assumption, DV-MDI-QKD is commonly
implemented by using independent laser sources for Alice and Bob. However, since the proof-of-principle demonstration in ref.~\cite{CVMDI} uses a common laser source for both of
them, this  
might open the door for side-channel attacks
on quantum state preparation. This is because at least one of the users has to encode
information on untrusted laser pulses accessible by Eve. 

Pirandola {\it et al.}~\cite{CVMDI} discussed
several potential solutions to the above problems in the Supplementary Information.
However, none of them has been implemented in ref.~\cite{CVMDI}.

\section{Addendum}

In this Appendix, we comment on a recent reply by Pirandola {\it et al.}~\cite{reply} to this paper. In particular, 

{\bf 1. Pirandola {\it et al.} agree with our main point.}

First of all, it is important to note that in their reply Pirandola {\it et al.}~\cite{reply} agree with our main point.
That is to say that, contrary to the statements in their Nature Photonics paper~\cite{CVMDI},
the secret key rate of DV-MDI-QKD with high-efficiency SPDs 
and good system alignment is sufficiently high for not only long distance communication but also metropolitan networks.
In fact, at metropolitan distances (say 20 km or more) in telecom fibers
(with a loss of 0.2 dB/km at 1550 nm wavelength), the key rate of DV-MDI-QKD
is only approximately two orders of magnitude away from the fundamental limit set by the TGW bound \cite{takeoka}.

The validity of this main point relies mainly on the existence of high-efficiency
SPDs at telecom wavelengths, which not only
have been used by a few research groups in the world in different recent experiments~\cite{SSPD,HDQKD,SSPDMDI,exp_det2,HDQKD2}, but also are already
commercially available \cite{company,company2}. These detector systems have
``fully-automated closed-cycle cryostats that deliver temperatures below 1K
without the consumption and running expense of liquid helium''~\cite{company}.

Most importantly, note that our main point holds independently of the performance of 
CV-MDI-QKD~\cite{CVMDI}. For completeness, however, below 
we address some other side points raised by Pirandola {\it et al.} in~\cite{reply}.  

{\bf 2. Experimental results of CV-MDI-QKD done in free-space, at a non-telecom wavelength, and 
using non-telecom detectors cannot and should not be used as a demonstration of telecom CV-MDI-QKD performance.}

In~\cite{CVMDI}, Pirandola {\it et al.} performed a {\it table-top free-space} experiment using a {\it single} laser at a wavelength
of 1064 nm that lies {\it far away from} the telecom band. Experimental results obtained 
under such conditions, however commendable, cannot be used to make {\it quantitative} statements 
about a realistic CV-MDI-QKD experiment that would be carried out {\it in fiber} using 
{\it two remotely located, independent, lasers} at a telecom wavelength (say 1550 nm) using 
telecom photodetectors. 

As will be discussed below in a point-by-point manner (see Points 5 \& 6), we find
such attempts to infer results from entirely different experimental platforms and conditions 
inappropriate and misleading. 

{\bf 3. CV-MDI-QKD could have an advantage over
DV-MDI-QKD only under rather restrictive conditions.}

We do not deny that CV-MDI-QKD might have an advantage over DV-MDI-QKD, but only in a
rather restrictive parameter space where a combination of assumptions/conditions
are simultaneously satisfied, namely, 
(i) asymptotic key rate for an infinitely long key;
(ii)  high-efficiency (well above 85\%) homodyne detectors,
(iii) highly asymmetric configuration where the relay is
close to one of the two users, Alice or Bob; and
(iv) low loss (i.e., short distance).

We encourage serious and careful future theoretical and experimental research in CV-MDI-QKD to address the above
relevant challenges. However, it is important to put things in perspective and
understand the nature of these relevant challenges now.

{\bf 4. For CV-QKD with coherent states, practical composable secure key rates 
against the most general
type of attacks have yet to be shown.}

It is important to clearly emphasise as well that the only relevant practical comparison
between DV-MDI-QKD and CV-MDI-QKD
is in the realistic finite-key setting scenario (not in the unrealistic infinitely long key situation).
In this respect, Pirandola {\it et al.}~\cite{reply} confuse the restricted class of {\it collective} attacks with the most general
class of {\it coherent} attacks. Indeed, CV-QKD can produce composable secure key rates against collective attacks
with a reasonable amount of signals distributed between Alice and Bob~\cite{Leverrier}. 
Unfortunately, however, current security proofs against the most general type of coherent attacks for CV-QKD with coherent states 
deliver basically zero key rate for any reasonable amount of 
signals~\cite{Leverrier}.
Indeed, as already noted in a recent
review on CV-QKD, composable security against coherent attacks ``has yet to be shown with coherent states'' 
(see line 2 of the first paragraph of p. 10 of~\cite{eleni}).

This strongly contrasts with DV-MDI-QKD, which can produce relatively high key rates
composable secure against 
general attacks~\cite{finite_mdiQKD}. Also, note that, in this scenario,
the minimum number
of signals that Alice and Bob need to distribute 
depends on the transmittance of the system. For instance,
it can be shown that when the 
detection efficiency of the relay is say $\eta=14.5\%$, the minimum number 
of signals is about $10^{10}$, while for $\eta=55\%$ ($\eta=93\%$) this number is $10^9$ ($10^8$) signals~\cite{finite_mdiQKD}.
In short, in reality, against the most general type of attacks,
current finite-key security proofs provide basically no key rate for CV-MDI-QKD but deliver a relatively high key 
for DV-MDI-QKD. 

This being said, from now on we relax this assumption and 
consider the asymptotic key rate in the limit of infinitely long keys. 

{\bf 5. The critical dependence of CV-MDI-QKD key rates on homodyne detection efficiency is down played by Pirandola {\it et al.}}

The performance of fiber-based CV-MDI-QKD depends crucially on the value of the detection efficiency of homodyne
detectors at telecom wavelengths. This is because, in an MDI-QKD setting, the relay cannot be assumed to be honest and its
detection efficiency can be fully exploited by Eve.
Therefore, the difference between $\eta=98\%$ and $\eta=80 \%$ can make or break the system.

\begin{figure*}[htb]
\centering
{\includegraphics [width=14.5cm]{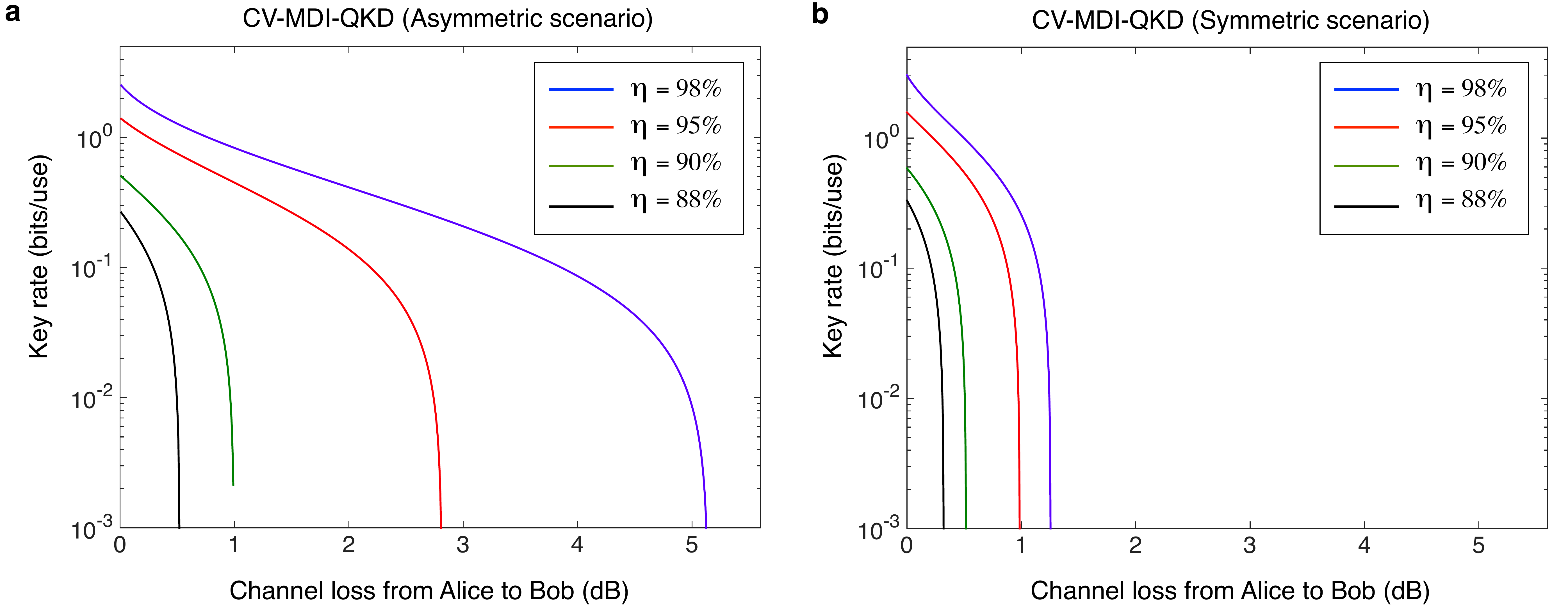}}
\caption{Secret key rate of CV-MDI-QKD in the asymptotic limit of infinitely long keys for different values of $\eta>85\%$. Like in Fig.~\ref{Fig}, we consider $\varepsilon\simeq0.01$, and a reconciliation efficiency $\xi=0.97$. 
\textbf{a},
Asymmetric scenario with a relay located close to Alice. \textbf{b}, Symmetric scenario with a relay placed in the middle of the users.
The blue lines correspond to the cases illustrated in Fig.~\ref{Fig} (i.e., $\eta=98\%$). 
 The key rate of CV-MDI-QKD is basically zero when $\eta< 85\%$ both for the asymmetric and symmetric configurations.
This figure highlights the dramatical performance decrease of CV-MDI-QKD even when $\eta$ is as high as $90\%$.} \label{Fig2}
\end{figure*}
For instance, the key rate 
(even in the asymptotic limit) of CV-MDI-QKD~\cite{CVMDI} is basically zero when $\eta<85\%$, 
while DV-MDI-QKD can in principle tolerate 
about 40 dB loss~\cite{MDIQKD}. 
Figs.~\ref{Fig2}a and~\ref{Fig2}b show the key rate of CV-MDI-QKD for several values of $\eta>85\%$.
These figures highlight the dramatical performance decrease of CV-MDI-QKD even when $\eta$ is as high as $90\%$. 
Pirandola {\it et al.}~\cite{CVMDI,reply} use $\eta=98\%$ for all the graphics that illustrate the key rate of CV-MDI-QKD.
For this, they argue that~\cite{reply} ``CV Bell detections routinely reach very high efficiencies, as typical
in many experiments~\cite{pir1,pir2,pir3,pir4} ($\eta=98\%$ in our setup)''. Unfortunately, neither the detectors
in their setup nor the ones in the references provided were fiber-coupled or in telecom wavelength. 

In addition, they claim~\cite{reply} that ``in a fibre-optic implementation of the protocol, coupling efficiency to a fibre can be as
high as 98-99\% and the quantum efficiency of photo-detectors can be $>$99\%, with an overall efficiency of about 97-98\%.''
Unfortunately again, Pirandola {\it et al.}~\cite{reply} seem to have ignored the difference between ``can be'' and ``is''. While there is no fundamental
limit for detectors to reach 97-98\% efficiency, 
most commercial off-the-shelf fiber-coupled detectors for the {\it telecom wavelength}
have responsivities less than 1 A/W. This number translates into 
a detection efficiency at 1550 nm of $\eta<80\%$, which is well below the minimum efficiency
of about 85\% required in CV-MDI-QKD.

{\bf 6. Pirandola {\it et al.}'s free-space experiment is not a properly designed QKD demonstration, thus the 
results shown in Fig.~\ref{Fig} are all ``purely-numerical''.}

In addition, one might want to point out that the experiment reported in~\cite{CVMDI}
is not a properly designed QKD demonstration, thus it is not appropriate to compare such
result with lab and field fiber-based experiments~\cite{timebinexp1,PolExp1,timebinexp2,PolExp2,MDIQKDexp2,MDIQKDexp3} 
in complete DV-MDI-QKD over practical distance at
all. This is so, because:

\begin{enumerate} 
\item {\it One single laser:} As already mentioned in the main text (see also Appendix~\ref{appD}), the experiment  in~\cite{CVMDI} uses one laser (with Alice and Bob very close to each other) to mimic two independent lasers, thus
bypassing the challenge of establishing a phase reference between two remote lasers, 
and therefore results in much less excess noise than would be expected had
they used two independent lasers. 

In this regard, Pirandola {\it et al.}~\cite{reply} claim that ``interfering signals from independent
laser sources is no longer a security issue or major experimental challenge in
CV-QKD'' due to the recent developments in~\cite{pir5,pir6}. While it is true 
that a potential solution to this problem has been proposed recently, 
note that this solution has not been implemented in~\cite{CVMDI} 
nor the additional expected phase noise has 
been considered in the simulations shown in refs.~\cite{CVMDI,reply}. 

\item {\it Table-top experiment at a wavelength outside the telecom band:} It is a table-top experiment 
conducted at 1064 nm through free-space. To use results obtained this way and plot the key rate
as a function of the distance (not loss) assuming 0.2 dB/km loss in fiber, as done in Fig.~5d~\cite{CVMDI} and in Fig.1~\cite{reply}, is entirely inappropriate. 
The loss in fiber at 1064 nm is approximately 1.5 dB/km. In fact, 1064 nm is shorter than the cut-off
wavelength of the standard telecom fiber, meaning that the fiber does not support single mode 
propagation at 1064 nm. 

\item {\it Incorrect noise model:}
Pirandola {\it et al.}~\cite{reply} claim that ``Finally, the excess noise in the experiment 
($\approx0.01$) is not low but typical (e.g., it is 0.008 in~\cite{CVQKD})." This is simply incorrect.

In CV-QKD, there are two major contributions to the overall excess noise: (i) the phase
noise in coherent detection (which can also include noise in state preparation), and (ii)
the electrical noise of the detector. In ref.~\cite{CVQKD}, the electrical noise
itself is already 0.015, almost twice as the value of 0.008 interpreted in~\cite{reply}.

The reasons to observe a so small excess noise ($\approx0.01$) in the experiment reported in~\cite{CVMDI} are, in our opinion, mainly two. First, 
the use of one single laser over a very short distance (see Point 6.1), 
thus the phase noise can be minimised. 
Second, the information provided in the supplementary material of~\cite{CVMDI} suggests
that the bandwidth of the homodyne detectors could be relatively small. In this condition, 
such results cannot be directly translated to high-speed CV-MDI-QKD. 

\end{enumerate}

In this regard, we would like to point out that the results shown in Fig.~\ref{Fig} are 
{\it all} ``purely-numerical''. In particular, 
we consider a quite favourable scenario for CV-MDI-QKD by assuming that: (i) a
reliable phase relation can be established between two remote lasers without
introducing any noise; and (ii) the loss of fiber coupling (required for fiber-based
QKD) or for a free-space optical receiver (required in free-space QKD) is zero. Both
assumptions are probably {\it over-optimistic}, so one might expect that the secure key rate of a
complete practical implementation of CV-MDI-QKD, when it is done, could be lower than 
that illustrated in Fig.~\ref{Fig}. In this sense, the lack of a CV-MDI-QKD demonstration that addresses 
all the points mentioned in this Appendix renders it difficult to make a ``{\it fair}'' comparison between both platforms. 

{\bf 7. CV-MDI-QKD could be suitable only for particular network architectures.}

Suppose a metropolitan network architecture with several users connected to a single relay (e.g., see Fig.~9 in ref.~\cite{network}). 
Due to the limited performance of symmetric CV-MDI-QKD, to guarantee that all users can communicate with each other (without 
trusting any user in the network), note that most of them (but one) should be located relatively close to the relay. 

{\bf 8. Final remarks.}

To conclude, we reply briefly to the statements that Pirandola {\it et al.}~\cite{reply} consider that are incorrect in our manuscript. In particular,

\begin{enumerate}

\item ``{\it The experimental rate of~\cite{CVMDI} is not an upper bound but a lower bound}''~\cite{reply}: Our manuscript does not claim 
that the {\it experimental} rate of~\cite{CVMDI} is an upper bound.

\item ``{\it The theoretical rate of~\cite{CVMDI} is optimal}''~\cite{reply}: Note that Fig.~\ref{Fig} 
already assumes that 
the {\it theoretical} rate of~\cite{CVMDI} is optimal. This being said, we admit that it was unclear to us whether or not the ``realistic Gaussian attack''
considered in~\cite{CVMDI} to derive the {\it theoretical} rate was indeed optimal, or if it provided an upper bound on the {\it theoretical} rate.
So, in our simulations we opted for the most favourable case for CV-MDI-QKD 
by considering that the attack is optimal (see Appendix~\ref{appB}). We are glad to see that Pirandola {\it et al.}~\cite{reply} confirm this. 

\item ``{\it Finite-size effects~\cite{pir7,CVQKD1,pir8} and composable security~\cite{Leverrier} support our experimental results}''~\cite{reply}: 
This statement is simply incorrect. See Point 4 above. 

\item ``{\it The relay doesn't need to be in AliceÕs lab}''~\cite{reply}: Our manuscript does not claim 
that the relay needs to be inside Alice's lab; in Fig.~\ref{Fig} we evaluate this case only because it is the 
most favourable one for CV-MDI-QKD. 

\item ``{\it Interfering signals from independent laser sources is no longer a security issue or major experimental challenge in CV-QKD}''~\cite{reply}: See 
Point 6 above. 

\item ``{\it Fast time-resolved homodyne detectors are available with bandwidths of 100MHz and more}''~\cite{reply}: 
None of the references provided by Pirandola {\it et al.}~\cite{reply} to support this statement are at telecom wavelengths, or have 
sufficiently high detection efficiency 
to deliver a significant key rate for CV-MDI-QKD. Also, note that our statement in the main text about 1 MHz CV-QKD simply reflects the
state-of-the-art of current clock rates for {\it complete} CV-QKD systems~\cite{CVQKD}, rather than the bandwidth of
homodyne detectors.

\end{enumerate}

\end{document}